\documentclass[12pt]{iopart}

\usepackage{amssymb}
\usepackage{amsfonts}
\usepackage{amstext}
\usepackage{graphicx}
\usepackage{amscd}

\newcommand{\ihbar}{\rmi \hbar}
\newcommand{\Te}{\mathbb{T}\rme}

\begin{document}

\title[Adiabatic population tracking for non-hermitian hamiltonians]{The role of the geometric phases in adiabatic populations tracking \\ for non-hermitian hamiltonians}

\author{A. Leclerc, D. Viennot and G. Jolicard}
\address{Institut UTINAM (CNRS UMR 6213, Universit\'e de Franche-Comt\'e, Observatoire de Besan\c con), 41bis Avenue de l'Observatoire, BP1615, 25010 Besan\c con cedex, France.}
\ead{Arnaud.Leclerc@utinam.cnrs.fr}

%\affiliation{Institut Carnot de Bourgogne (CNRS UMR 5209, Universit\'e de Bourgogne), BP 47870, 21078 Dijon, France.}
%\author{Pierre Joubert}
%\affiliation{Institut UTINAM (CNRS UMR 6213, Universit\'e de Franche-Comt\'e), 16 route de Gray, 25000 Besan\c con cedex, France.}
%\author{John P. Killingbeck}
%\affiliation{Centre for Mathematics, University of Hull, Hull HU6 7RX, UK.}
 
\begin{abstract}
The definition of instantaneous eigenstate populations for a dynamical non-self-adjoint system is not obvious. The na\"ive direct extension of the definition used for the self-adjoint case leads to inconsistencies; the resulting artifacts can induce a false inversion of population or a false adiabaticity. We show that the inconsistency can be avoided by introducing geometric phases in another possible definition of populations. An example is given which demonstrates both the anomalous effects and their removal by our approach.
\end{abstract}

%\ams{}
\submitto{\JPA}

%\maketitle

\section{Introduction}
The adiabatic approximation is currently a commonly used tool to study quantum dynamical systems \cite{messiah}. Let $s=t/T$ be the reduced time ($T$ being the total duration which is assumed to be very large). Let $|a,x(s) \rangle$ be an instantaneous eigenvector of a selfadjoint hamiltonian $H(x(s))$ associated with a non degenerate eigenvalue $E_a$, i.e.
\begin{equation}
H(x(s))|a,x(s)\rangle=E_a |a,x(s)\rangle.
\end{equation}
$H$ depends on the reduced times $s$ from some classical parameters $x$. Starting from the initial state $\psi(0) = |a,x(0) \rangle$, 
%%%%%%%%%%%%%%%%%%%%%%%%%%%%%
%RAJOUT
the state vector propagation obeys the time-dependent Schr\"odinger equation
\begin{equation}
 \frac{\rmi}{T} \frac{\partial \psi(sT)}{\partial s} = H(x(s)) \psi(sT).
\end{equation}
%%%%%%%%%%%%%%%%%%%%%%%%%%%%%
The adiabatic theorem \cite{messiah,nenciu} states that the systems continuously follows the same eigenstate $|a,x(s) \rangle$ if the hamiltonian is gradually modified; that is
\begin{equation}
\psi(sT) = c_a(s) |a,x(s) \rangle + \sum_{b \not= a} c_b(s)|b,x(s) \rangle
\end{equation}
with
\begin{eqnarray}
\forall b\not=a, \left|\frac{c_b}{c_a} \right| \ll 1 ,
\end{eqnarray}
$|c_a(s)|^2$ being called the population of the state $a$, if the following adiabatic criterion is fulfilled:
\begin{equation}
\forall b \not=a, \quad |\langle b,x(s)|\frac{\rmd}{\rmd s}|a,x(s)\rangle| = \left|\frac{\langle b,x(s)| \frac{\rmd H}{\rmd s} |a,x(s) \rangle}{E_a(x(s)) - E_b(x(s))} \right| \ll 1
\end{equation}
which requires satisfaction of the gap condition: $\exists g>0, \forall s \quad |E_a(x(s)) - E_b(x(s))| > g$, and also slow time variations within the hamiltonian $H(x(s))$. This leads to the wavefunction
\begin{equation}
\label{psiadiab}
\fl \psi(sT) \simeq \exp \left({- \ihbar^{-1} T \int_0^s E_a(x(s'))\rmd s'}\right) \exp \left({- \int_0^s A(x(s'))\rmd s'}\right) |a,x(s) \rangle ,
\end{equation}
where $- \ihbar^{-1} T \int_0^s E_a(x(s))\rmd s$ is the dynamic phase and $- \int_0^s A(x(s))\rmd s$ is the geometric phase. The geometric phase generator is defined as
\begin{equation}
A(x)=\langle a,x | \rmd | a,x \rangle \Rightarrow A(x(s)) \rmd s = \langle a,x(s)|\frac{\rmd}{\rmd s}|a,x(s)\rangle \rmd s.
\end{equation}

Studies on adiabaticity and geometric phases can now be usefully extended to dissipative systems which use non-selfadjoint hamiltonians to describe resonance phenomena \cite{moiseyev,mailybaev,viennot,viennot2,dridi}.
%ajout de David suite referee ici
When resonances are defined by some scattering boundary conditions the resonance states are unbounded. This means that numerical and theoretical treatment of the resonances is a difficult task.  Spectral deformation methods are commonly used to solve this problem. The first one, the complex scaling method \cite{moiseyev2}, effectively transforms the resonance states into bound states. The hamiltonian becomes non-selfadjoint, and the resonances calculations produce non-real eigenvalues and the associated eigenvectors. The real part of a resonance eigenvalue corresponds to the resonance energy and its imaginary part corresponds to the resonance width (the inverse of the resonance lifetime). A second approach is the optical potential method \cite{jolicard}, which has similar properties. %but has the supplementary advantage of permitting a description of the resonance phenomenon using a finite size configuration space.
These two methods of resonance modelling involve non-selfadjoint hamiltonians and are used successfully to treat molecular photodissociation problems \cite{he,ben,atabek}. The present paper deals with non-hermitian matrices. Such matrices can be viewed as effective hamiltonians associated with the hamiltonians of resonance problems. Effective hamiltonian techniques use small hamiltonians to describe complex systems without loss of informations. The principal effective hamiltonians techniques are the partitioning technique \cite{reviewgeorges1}, the quantum KAM method \cite{reviewguerin}, the adiabatic elimination method \cite{reviewguerin} and the Bloch wave operator method \cite{reviewgeorges1}. Small non-selfadjoint matrices have also been used as hamiltonians to describe some photoionization phenomena modelled by bound states coupled with a structureless continuum (these states are then viewed as resonances) \cite{kylstra,magunov,gryzlova}.
Finally we should note that the formalism used in this paper can be used to treat open quantum systems \cite{sarandy,yi} which are described by a Lindblad equation
\begin{equation*}
\rmi \hbar \frac{\partial \rho}{\partial t} = L(\rho) 
\end{equation*}
where $\rho$ is a density matrix (a traceless positive selfadjoint matrix) and the Lindbladian $L$ is a "superoperator". For a finite dimensional Hilbert space, $\dim \mathcal H = n$, the space of the density matrices can be identified with an $n^2$-dimensional Hilbert space $\mathcal L$ (usually called the Liouville space). In the Liouville space, the Lindblad equation takes the form of the usual Schr\"odinger equation. If the Lindbladian is reduced to be a commutator with a selfadjoint hamiltonian, i.e. $L(\rho) = [H,\rho]$ then $L$ is a selfadjoint operator in the Liouville space $\mathcal L$. Usually, however, the Lindbladian is $L(\rho) = [H,\rho] - \frac{\rmi}{2} \sum_k (\{\Gamma_k^\dagger \Gamma_k,\rho\} + 2 \Gamma_k \rho \Gamma_k^\dagger)$ where the operators $\Gamma_k$ model the different decohering processes ($\{.,.\}$ is the anticommutator). In this case, $L$ is a non-selfadjoint operator in $\mathcal L$. The non-real eigenvalues of $L$ (its resonances) are associated with the decoherence. The present work can be applied to a such system, but for the sake of simplicity we will use only the name "hamiltonian" even though it could represent a Lindbladian in the Liouville space. 
%%%

For non-selfadjoint hamiltonians, the adiabatic theorem \cite{nenciurasche,joye} involves a criterion on the dissipation rate of the quantum system studied (i.e. on the imaginary part of the instantaneous eigenvalue considered). 
Naively, we could imagine that the adiabatic approximation for dissipative quantum systems should take the form  $\psi(sT) = \sum_b c_b(s) |b, x(s)\rangle$ with $\forall b \not=a$, $\left|\frac{c_b}{c_a} \right|  \ll 1$. The following section shows that this approximation is inconsistent, since the populations $|c_b|^2$ are not well defined. Consequently a better definition of the instantaneous populations is proposed
%%%%%%%%%%%%%%%%%%%%%%%%%
% RAJOUT
in section~\ref{sectiondefinition}.
This definition is equivalent to the c-product condition (described in \cite{moiseyev}) for the case of symmetric non-hermitian hamiltonians and remains valid even for the non-symmetric case.
%%%%%%%%%%%%%%%%%%%%%%%%%
The illustrative example of a 2-state dissipative system is then described in section \ref{example},
%%%%%%%%%%%%%%%%%%%%%%%%%
% RAJOUT
with an application to an adiabatic state flip \cite{uzdin,berry} which is generated by following a loop around an exceptional point in the parameter plane.
%%%%%%%%%%%%%%%%%%%%%%%%%

\section{Populations of the instantaneous eigenvectors \label{sectiondefinition}}

For the sake of simplicity we consider a non-selfadjoint hamiltonian $H(x)$ of rank 2 (the discussion can easily be generalized to higher dimensions). 
%Here we use neither the c-product formalism \cite{gilary}, nor the rigged Hilbert space formalism \cite{bohm,madrid}, both of which are often used in the analysis of non-hermitian quantum problems.
%Here we do not use the c-product formalism \cite{gilary} which is often used in the analysis of non-hermitian quantum problems.  
We deal with a biorthogonal basis set and we %cautiously
use the standard scalar product. Let $\{E_1(x),E_2(x)\}$ be the two instantaneous eigenvalues of $H(x)$ (assumed to be diagonalizable) and $\{|1,x\rangle,|2,x\rangle\}$ the two associated eigenvectors, and let $\{|1*,x\rangle,|2*,x\rangle\}$ be the biorthogonal basis, that is:
\begin{eqnarray}
H(x) |a,x\rangle &=& E_a (x) | a ,x\rangle, \\
H(x)^\dagger |a*,x\rangle &=& \overline{E_a(x)}|a*,x\rangle,
\end{eqnarray}
\begin{equation}
\langle a*,x|b,x \rangle = \delta_{ab} .
\end{equation}
(The bar denotes the complex conjugate).
We note that in certain highly symmetric cases (i.e. when $H^t=H$)
the $\{|a,x\rangle\} $ and the $\{|a*,x\rangle\}$ are related by a simple complex conjugation rule (formalized by the c-product \cite{gilary,moiseyev}). \\
Let 
\begin{equation}
\label{A}
A(x)  =  \left(\begin{array}{cc} \langle 1*,x|\rmd|1,x\rangle & \langle 1*,x|\rmd|2,x\rangle \\ \langle 2*,x|\rmd|1,x\rangle & \langle 2*,x|\rmd|2,x\rangle \end{array}\right)
\end{equation}
be the matrix of the geometric phase generators and of the non-adiabatic coupling. Let
\begin{equation}
\psi(t)  =  U(t,0) |1,x(s=0) \rangle
\end{equation}
be the wave function for an evolution $s \mapsto x(s)$ ($U(t,0)$ being the evolution operator). As in conservative systems, we could write
\begin{equation}
\psi(sT) = c_1(s) |1,x(s) \rangle + c_2(s)|2,x(s) \rangle ,
\end{equation}
$|c_1(s)|^2$ being the ``population'' of the state 1 and $|c_2(s)|^2$ the ``population'' of the state~2;
the adiabatic approximation corresponds to the case where $|c_2|^2$ is negligible. However this definition presents a significant problem. By constrast with the conservative case, where we choose an orthonormal eigenbasis to represent the dynamics, the normalization convention is arbitrary for a biorthogonal basis. Indeed, if $\left(\{|1,x\rangle,|2,x\rangle\},\{|1*,x \rangle,|2*,x\rangle\} \right)$ constitutes a biorthogonal eigenvectors system, this is also the case for $\left(\{\widetilde{|1,x\rangle},\widetilde{|2,x\rangle}\},\{\widetilde{|1*,x \rangle},\widetilde{|2*,x\rangle}\} \right)$ with
\begin{eqnarray}
\widetilde{|a,x\rangle} & = & \lambda_a(x) |a,x \rangle \\
\widetilde{|a*,x\rangle} & = & (\overline{\lambda_a(x)})^{-1} |a*,x \rangle
\end{eqnarray}
for all arbitrary functions $\lambda_a(x) \in \mathbb C^*$. We must then set
\begin{equation}
\psi(sT) = \tilde c_1(s) \widetilde{|1,x(s) \rangle} + \tilde c_2(s)\widetilde{|2,x(s) \rangle}
\end{equation}
with
\begin{equation}
\tilde c_a(s) = \frac{c_a(s)}{\lambda_a(x(s))}.
\end{equation}
The population $|\tilde c_a|^2 = \frac{|c_a|^2}{|\lambda_a|^2}$ then depends on the arbitrary normalization convention. The calculation of the instantaneous populations then makes no sense, since the adiabatic multiplier $c_1$ can artificially grow if the normalization of $|a,s\rangle$ decreases.

To solve this problem, we introduce a definition of the instantaneous populations using geometric phases. We set
\begin{eqnarray}
\psi(sT) &=& d_1(s) \exp \left(- \int_0^s A_{11}(x(s'))\rmd s'\right) |1,x(s) \rangle \nonumber \\
&&+ d_2(s) \exp \left({- \int_0^s A_{22}(x(s'))\rmd s'}\right) |2,x(s) \rangle
\label{gooddefinition}
\end{eqnarray}
with $\langle a,x(s=0)|a,x(s=0)\rangle=1$.
The instantaneous populations are now defined by $|d_a(s)|^2$. 
Immediately we notice that for the conservative case this definition coincides with the standard definition of the instantaneous populations since $|\rme^{-\int_0^s A_{aa}(x(s'))\rmd s'}|^2 =1$ in a self-adjoint system ($A_{aa}$ is purely imaginary; the standard definition of the instantaneous populations is invariant under arbitrary changes to the phase convention of the eigenvectors). By making an arbitrary change in the normalization convention, we have
\begin{eqnarray}
\tilde A_{aa}(x) & = & \widetilde{\langle a*,x|} \rmd \widetilde{|a,x\rangle} \\
& = & \frac{\rmd \lambda_a(x)}{\lambda_a(x)} + \langle a*,x| \rmd |a,x\rangle \\
& = & \rmd \ln \lambda_a(x)  + A_{aa}(x).
\end{eqnarray}
We then have
\begin{equation}
\rme^{-\int_0^s \tilde A_{aa}(x(s'))\rmd s'} = \frac{\lambda_a(x(0))}{\lambda_a(x(s))} \; \rme^{-\int_0^s A_{aa}(x(s'))\rmd s'}.
\end{equation}
In order to preserve the initial condition, we set $\lambda_a(x(0)) = 1$. This leads to
 \begin{equation}
\fl \psi(sT)  =  d_1(s) \rme^{- \int_0^s \tilde A_{11}(x(s'))\rmd s'} \widetilde{|1,x(s) \rangle}  + d_2(s) \rme^{- \int_0^s \tilde A_{22}(x(s'))\rmd s'} \widetilde{|2,x(s) \rangle}.
\end{equation}
The definition of the instantaneous populations, $|d_a|^2$, is now invariant even in the event of arbitrary changes in the normalization convention. Here we insist on the fact that the coefficients $d_a$ are more intrinsic than the coefficients $c_a$.\\
Owing to dissipation the total population strays away from $1$ (except for $s=0$). We have $1-(|d_1|^2 + |d_2|^2) \geq 0$, but this is not a properly defined dissipation rate (i.e. $|d_1|^2 + |d_2|^2 \not= \|\psi\|^2$). The dissipation rate is given by $(1-\|\psi(sT)\|^2)$ with
\begin{eqnarray}
\|\psi(sT)\|^2 & = & \sum_{a,b} \overline{d_b(s)}d_a(s) \exp \left({-\int_0^s (A_{aa}(x(s'))+\overline{A_{bb}(x(s'))})\rmd s'}\right) \nonumber \\ 
& & \qquad \times \langle b,x(s)|a,x(s) \rangle \\
& = & \sum_{a,b} \overline{d_b(s)} \eta_{ba}(s) d_a(s) \\
& = & D(s)^\dagger \eta(s) D(s)
\end{eqnarray}
with $D(s) = \left( \begin{array}{c} d_1(s) \\ d_2(s) \end{array} \right)$ and $\eta_{ba}(s) = \rme^{-\int_0^s (A_{aa}(x(s'))+\overline{A_{bb}(x(s'))})\rmd s'} \langle b,x(s)|a,x(s) \rangle$. $\eta$ constitutes an $s$-dependent scalar product for $D(s)$, the representation of the wave functions in the instantaneous (non-orthonormal) eigenbasis. We note that $\eta$ is well defined since it is independent of the normalization convention, $\tilde \eta = \eta$. The scalar product matrix satisfies the following differential equation:
\begin{equation}
\frac{\rmd\eta}{\rmd s} = \hat A^\dagger \eta + \eta \hat A
\end{equation}
with 
\begin{equation}
\hat A_{ab}(s) = \langle a*,x(s)| \rme^{\int_0^s A_{aa}(x(s'))\rmd s'} \frac{\rmd}{\rmd s} \left( \rme^{-\int_0^s A_{bb}(x(s'))\rmd s'}|b,x(s) \rangle \right). 
\end{equation}
We can then write
\begin{equation}
\eta(s) = \left(\Te^{\int_0^s \hat A(s')\rmd s'} \right)^\dagger \eta(0) \Te^{\int_0^s \hat A(s')\rmd s'}
\end{equation}
where $\Te$ is the $s$-ordered exponential (the Dyson series).\\
Finally, if we want the instantaneous total population to be consistent with the dissipation rate, then the instantaneous population must be defined as $\alpha(s) |d_a(s)|^2$ with~$\alpha = \frac{D^\dagger \eta D}{D^\dagger D}$.\\

Using this analysis, we propose that the consistent adiabatic approximation is
\begin{equation}
\left|\frac{d_2}{d_1}\right| \ll 1
\end{equation}
and not $\left|\frac{c_2}{c_1}\right| \ll 1$. In a similar manner, the adiabatic criterion must be independent of the arbitrary normalization convention:
\begin{eqnarray}
& & \rme^{\int_0^s \Re\mathrm e(A_{bb}(x(s')) -A_{aa}(x(s')))\rmd s'} |\langle b*,x(s)|\frac{\rmd}{\rmd s}|a,x(s)\rangle|\nonumber \\
& &  = \rme^{\int_0^s \Re\mathrm e(A_{bb}(x(s')) -A_{aa}(x(s')))\rmd s'} \left|\frac{\langle b*,x(s)| \frac{\rmd H}{\rmd s} |a,x(s) \rangle}{E_a(x(s)) - E_b(x(s))} \right|\nonumber \\
& & \qquad  \ll 1.
\end{eqnarray}

We should point out that the use of the parallel transport condition, i.e. the definition of the populations with eigenvectors $\{\widehat{|a,x(s)\rangle}\}_a$ and $\{\widehat{|a*,x(s)\rangle}\}_a$ such that
\begin{equation}
\widehat{\langle a*,x(s)|} \frac{\rmd}{\rmd s} \widehat{|a,x(s)\rangle} = 0
\end{equation}
implicitely takes into account the instantaneous population definition which includes the geometric phases. Indeed, it is easy to show that the $\{\widehat{|a,x(s)\rangle}\}_a$ are related to an arbitrary set of eigenvectors $\{|a,x\rangle\}_a$ by
\begin{equation}
\widehat{|a,x(s)\rangle} = \rme^{- \int_0^s A_{aa}(x(s'))\rmd s'}|a,x(s) \rangle.
\end{equation}
%%%%%%%%%%%%%%%%%%%%%%%%%%
% RAJOUT
In the particular case of a symmetric matrix, the left eigenvector are the complex conjugates of the right eigenvectors, so that the parallel transport condition is also equivalent to the c-product normalization condition for the eigenvectors ${\vert a,x(s)_{\text{c.p.}} \rangle}$, fixed by \cite{moiseyev}
\begin{equation}
\overline{ \langle a,x(s)_{\text{c.p.}}} \vert a,x(s)_{\text{c.p.}} \rangle = 1.
\label{cproduct}
\end{equation}
%
%%%%%%%%%%%%%%%%%%%%%%%%%%
The result of this paper could be %then 
expressed without explicit reference to the geometric phases by saying that the treatment of the population tracking for non-hermitian systems necessary needs to impose the parallel transport condition (whereas this is not necessary for hermitian systems). Nevertheless, we prefer here to make the geometric phases appear explicitely. Indeed the $\{\widehat{|a,x(s)\rangle}\}_a$ are defined only along the path in the parameter space defined by $s \mapsto x(s)$; for a different path $s \mapsto x'(s)$ the eigenvectors $\{\widehat{|a,x(s)\rangle'}\}_a$ are different. It is then more general to consider $\{|a,x\rangle\}_a$ (without a parallel transport condition) which are defined globally on the whole of the parameter space. \\
Moreover we note that we cannot use the parallel transport condition to define the eigenvectors if the path is closed $x(s=0)=x(s=1)$, because of the double definition of the eigenvectors at $x(0)=x(1)$:
\begin{equation}
\widehat{|a,x(1)\rangle} = \rme^{- \int_0^1 A_{aa}(x(s))\rmd s} |a,x(1) \rangle \not= |a,x(0)\rangle = \widehat{|a,x(0) \rangle}.
\end{equation}
The impossibility of giving a single definition of the eigenvectors in the parallel transport condition $\rme^{- \int_0^1 A_{aa}(x(s))\rmd s}$ is called the holonomy of the parallel transport.\\

The next section illustrates that numerical artifacts due to bad definitions of the populations can induce a false adiabaticity or a false inversion of population in numerical simulations.

\section{Illustrative example \label{example}}

%%%%%%%%%%%%%%%%%%%%%%%%%%%
% RAJOUT
%
The definition (\ref{gooddefinition}) is relevant for non-hermitian symmetric or non-symmetric matrices.
While the formalism used here should work for larger more realistic hamiltonians and for Floquet-type hamiltonians,
we study here a low dimensional illustrative example for which the data can be shown conveniently.
%%%%%%%%%%%%%%%%%%%%%%%%%%%
Let the parameter dependent hamiltonian be
\begin{equation}
H(w,z) = \left(\begin{array}{cc} 0 & w \\ \bar w & 2z \end{array}\right) = \left(\begin{array}{cc} 0 & \Omega e^{\rmi \phi} \\ \Omega e^{- \rmi \phi} & 2 \Delta - \rmi \frac{\Gamma}{2} \end{array} \right)
\label{matrice}
\end{equation}
$(w,z) \in \mathbb C^2$. 
%%%%%%%%%%%%%%%%%%%%%%
%RAJOUT EVENTUEL
%
%In the framework of the rotating wave approximation,
%%%%%%%%%%%%%%%%%%%%%%
This hamiltonian is associated with a quantum bound state coupled to a quantum resonance (with resonance width $\Gamma$) by a laser field with amplitude $\Omega$ and phase $\phi$. The laser field is quasiresonant with the transition from the bound state to the resonance with a detuning value equal to $\Delta$. We assume that we can modulate the complex numbers $(w,z)$ to generate a dynamics.\\

%RAJOUT DU TITRE %%%%%%%%%%%%%%%%%%%%%%
\subsection{False adiabaticity}

\label{falseadiabaticitysubsection}

The spectrum of $H$ is
\begin{eqnarray}
E_{1}(w,z) & = & z - \sqrt{|w|^2+z^2} = z - v\\
E_{2}(w,z) & = & z + \sqrt{|w|^2+z^2} = z + v
\end{eqnarray}
where $v = \sqrt{|w|^2+z^2} = z \sqrt{1+\frac{|w|^2}{z^2}}$ (we choose the Riemann sheet such that $\sqrt{z^2}=z$). We restrict our attention to the parameters $(w,z)$ such that $\Im\mathrm m(E_2-E_1) = 2 \Im\mathrm m(v)<0$ ($E_1$ is the less dissipative).\\
We can easily verify that the eigenvectors of $H(w,z)$ are
%\begin{subequations}
\begin{eqnarray}
|1,w,z \rangle & = & \gamma_{1}\left(\begin{array}{c} z + v \\ - \bar w \end{array} \right) \\
|2,w,z \rangle & = & \gamma_{2}\left(\begin{array}{c} z-v \\ - \bar w \end{array} \right)
\end{eqnarray}
%\end{subequations}
where $\gamma_{a}$ are the appropriate factors to fix the initial norm of the basis vectors to $1$, i.e.
%\begin{subequations}
\begin{eqnarray}
 \gamma_{1}&=&\frac{1}{\sqrt{(|z(0)+v(0)|^2+|w(0)|^2)}}\\
\gamma_{2}&=&\frac{1}{\sqrt{(|z(0)-v(0)|^2+|w(0)|^2)}}.
\end{eqnarray}
%\end{subequations}
The biorthogonal basis set is
%\begin{subequations}
\begin{eqnarray}
|1*,w,z \rangle & = & \frac{1}{2\bar v(\bar v+ \bar z)\gamma_{1}^2} |1,w,\bar z \rangle \\
|2*,w,z \rangle & = & \frac{1}{2\bar v(\bar v- \bar z)\gamma_{2}^2} |2,w,\bar z \rangle.
\end{eqnarray}
%\end{subequations}
The associated geometric phase generators are
%\begin{subequations}
\begin{eqnarray}
A_{11} & = & \frac{\bar w\dot w + w\dot{\bar w}}{4v^2}+\frac{w\dot{\bar w}}{2v(v+z)}+\frac{(z+v)\dot z}{2v^2}  \\
A_{22} & = & \frac{\bar w\dot w + w\dot{\bar w}}{4v^2}+\frac{w\dot{\bar w}}{2v(v-z)}-\frac{(v-z)\dot z}{2v^2}.
\end{eqnarray}
%\end{subequations}
%We can note that the norm of the eigenstate 1 ten\rmd s to zero at the limit $|w|\to 0$ whereas this is not the case for the norm of the eigenstate 2.
%\begin{eqnarray}
%\lambda_1(w,z) & = & \| |1,w,z \rangle \| \nonumber \\
%& = & \sqrt{|z|^2 \left|1-\sqrt{1+\frac{|w|^2}{z^2}}\right|^2 + |w|^2} \nonumber \\
%\label{lambda1}
%& & \qquad \xrightarrow[|w|\to 0]{} 0 \\
%\lambda_2(w,z) & = & \| |2,w,z \rangle \| \nonumber \\
%& = & \sqrt{|z|^2 \left|1+\sqrt{1+\frac{|w|^2}{z^2}}\right|^2 + |w|^2} \nonumber \\
%\label{lambda2}
%& & \qquad \xrightarrow[|w|\to 0]{} 2|z|
%\end{eqnarray}

We can verify that another possible eigenstate 1 reads
\begin{equation}
\widetilde{|1,w,z \rangle} = \frac{\gamma_{1}}{\beta} \left(\begin{array}{c} w \\ z-v \end{array} \right)
\end{equation}
with
\begin{equation}
|1,w,z \rangle = \beta \;\frac{v+z}{w} \widetilde{|1,w,z \rangle}
\end{equation}
and
\begin{equation}
\beta=\frac{w(0)}{v(0)+z(0)}.
\end{equation}
The factor $\beta$ ensures that $|1,w(0),z(0)\rangle=\widetilde{|1,w(0),z(0) \rangle}$ in order
to preserve the initial condition. We assume that $\beta \neq 0$.
The associated biorthogonal eigenvector is
\begin{equation}
 \widetilde{|1*,w,z\rangle}=\frac{|\beta|^2}{2\bar v(\bar v- \bar z)\gamma_{1}^2} \widetilde{|1,w,\bar z \rangle}.
\end{equation}
The normalization factor between the two conventions approaches zero at the limit $|w| \to 0$,
\begin{equation}
\label{zvw}
\lim_{|w|\to 0} \beta \frac{w}{v+z} = 0.
%\beta \frac{w}{v+z} \xrightarrow[|w|\to 0]{}  0.
%
\end{equation}
With this normalization, the first geometric phase generator becomes
\begin{equation}
\tilde A_{11}  = \frac{\bar w\dot w + w\dot{\bar w}}{4v^2}+\frac{\bar w\dot w}{2v(v-z)}-\frac{(v-z)\dot z}{2v^2}.
\end{equation}

These properties induce the following analysis. Let 
\begin{equation}
\fl \psi(sT) = d_1(s)\rme^{-\int_0^s \tilde A_{11}(s')\rmd s'} \widetilde{|1,w(s),z(s) \rangle}+  d_2(s)\rme^{-\int_0^s A_{22}(s')\rmd s'} |2,w(s),z(s) \rangle \label{psi}
\end{equation}
 be the solution to the Schr\"odinger equation for the evolution $s \mapsto (w(s),z(s))$ with the initial state $\psi(0) = |1,w(0),z(0) \rangle$. We suppose that the adiabatic approximation is valid, i.e. $\left| \frac{d_2}{d_1} \right| \ll 1$. If we consider the naive definition of the population by setting 
\begin{equation*}
\psi(sT) =  c_1(s)|1,w(s),z(s) \rangle +  c_2(s) |2,w(s),z(s) \rangle 
\end{equation*}
then the quotient
\begin{eqnarray}
\left| \frac{ c_1}{c_2} \right| & = & \frac{1}{\beta} \frac{|w|}{|z+v|} \left| \frac{\tilde c_1}{c_2} \right| \\
& = & \frac{1}{\beta} \frac{|w|}{|z+v|} \left| \frac{d_1}{d_2} \right| \rme^{\int_0^s \Re\mathrm e(A_{22}(s')-\tilde A_{11}(s'))\rmd s'}
\end{eqnarray}
with $\psi(sT) = \tilde c_1(s) \widetilde{|1,w(s),z(s) \rangle} + c_2(s) |2,w(s),z(s) \rangle$. We note that if $\Im \mathrm m (w) = 0$ then $\Re\mathrm e(A_{22}-\tilde A_{11}) = 0$ and $\left| \frac{c_1}{c_2} \right| = \frac{|w|}{|\beta||z+v|} \left| \frac{d_1}{d_2} \right|$. The limit (\ref{zvw}) can induce
\begin{equation}
|w| \ll 1 \Rightarrow \left| \frac{d_2}{d_1} \right| \ll 1 \text{ and }  \left| \frac{c_1}{c_2} \right| \ll 1.
\end{equation}
Thus we observe a false non-adiabaticity due to the badly defined populations. In this case we also have a false population inversion (in the sense that population 1 is negligible with respect to population 2 using the badly defined populations, while the well-defined populations give the inverse result).\\

Conversely, let $\psi(sT)$ [cf. (\ref{psi})] % = d_1(s)e^{-\int_0^s \tilde A_{11}(s')\rmd s'} \widetilde{|1,w(s),z(s) \rangle} + d_2(s)e^{-\int_0^s A_{22}(s')\rmd s'} |2,w(s),z(s) \rangle$ 
be the solution to the Schr\"odinger equation for an evolution starting with an initial state $\psi(0) = |2,w(0),z(0) \rangle$ such that the adiabatic approximation is not satisfied, i.e. for $s$ sufficiently large, $|d_1| \sim |d_2|$. By the same arguments, if $|w| \ll 1$ we have 
\begin{equation}
\left| \frac{d_1}{d_2} \right| \sim 1 \text{ and } \left| \frac{c_1}{c_2} \right| \ll 1.
\end{equation}
In this case, we obtain a false adiabaticity due to the badly defined populations.

\begin{figure}[!ht]
\centering
 \includegraphics[width=0.6\linewidth]{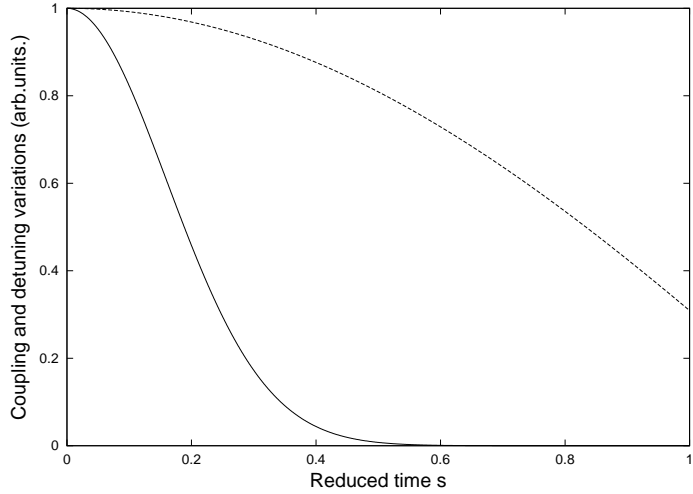}
\caption{Time variations of the Rabi frequency $w$ (line) and of the detuning $\Re\mathrm e{(z)}$ (dashes).}
\label{fig1}
\end{figure}

\begin{figure}[htp]
\centering
 \includegraphics[width=0.6\linewidth]{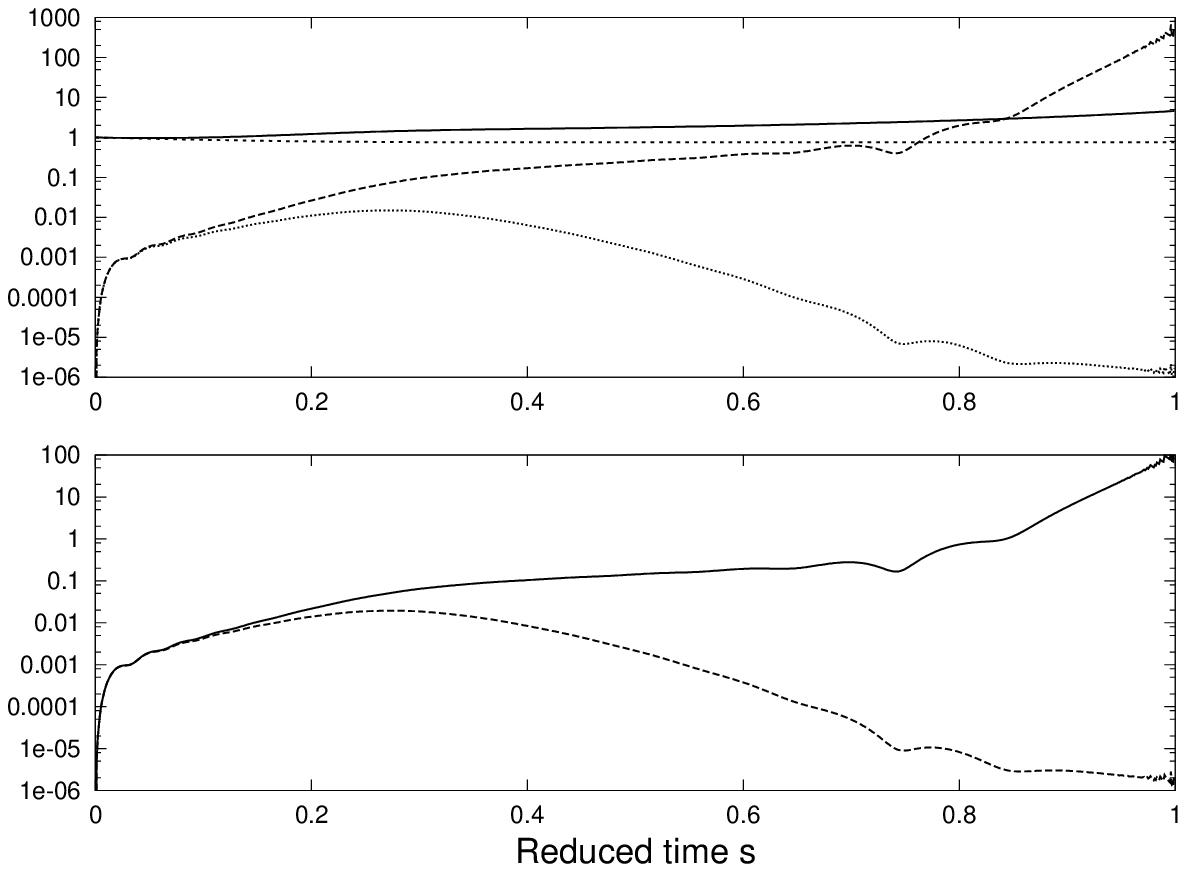}
\caption{Top: $|c_1(s)|$ (line), $|c_2(s)|$ (long dashes), $|d_1(s)|$ (short dashes), $|d_2(s)|$ (dots); bottom:
ratios $|c_2(s)/c_1(s)|$ (line) and $|d_2(s)/d_1(s)|$ (dashes). The initial state is $|1,w(0),z(0)\rangle$. $\Gamma=0.1$.}
\label{fake_inversion}
\end{figure}

\begin{figure}[htp]
\centering
 \includegraphics[width=0.6\linewidth]{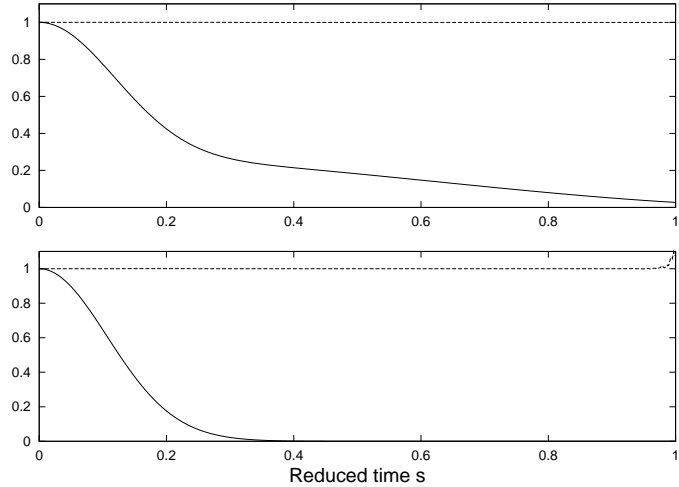}
\caption{Top: $\langle 1,w,z|1,w,z\rangle$ (line), $\langle 1,w,z|1,w,z\rangle\times |\rme^{\int_0^s A_{11}(s')\rmd s'}|^2$ (dashed line). Bottom: $\langle 2,w,z|2,w,z\rangle$ (line), $\langle 2,w,z|2,w,z\rangle\times |\rme^{\int_0^s A_{22}(s')\rmd s'}|^2$ (dashed line). This corresponds to the case of a fake inversion of population with $\Gamma=0.1$.}
\label{normes}
\end{figure}

\begin{figure}[htp]
\centering
 \includegraphics[width=0.6\linewidth]{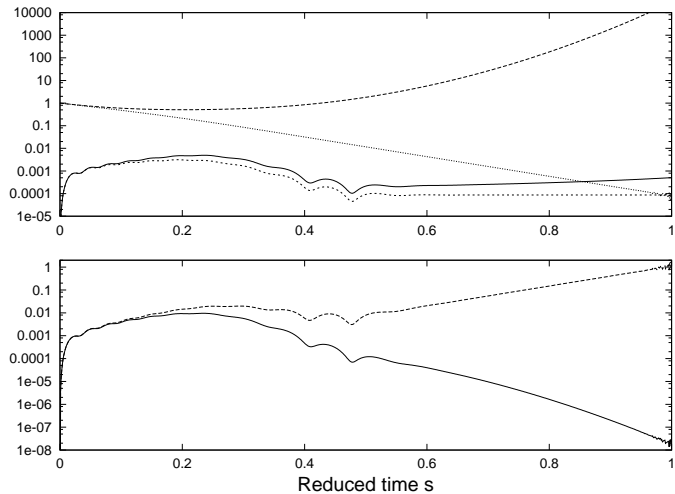}
\caption{Top: $|c_1(s)|$ (line), $|c_2(s)|$ (long dashes), $|d_1(s)|$ (short dashes), $|d_2(s)|$ (dots); bottom:
ratios $|c_1(s)/c_2(s)|$ (line) and $|d_1(s)/d_2(s)|$ (dashes). The initial state was $|2,w(0),z(0)\rangle$. $\Gamma=0.2$.}
\label{fake_adiabaticity}
\end{figure}

Figure \ref{fig1} shows a simple example of a gaussian variation for $w(s)$ (coupling) and a decreasing cosine function for $\Re\mathrm e{(z)}$ (detuning), so that $|w(t)|\simeq0$ near the end of the considered time interval $[0,100]$.
The chosen functions are
%\begin{subequations}
 \begin{eqnarray}
  w(s)&=&w_0 e^{-\frac{s^2}{2\sigma^2}}\\
\Delta(s)&=&\Re\mathrm e(z(s))= \Delta_0 \cos(0.4 \pi s)
 \end{eqnarray}
%\end{subequations}
with $w_0=1$, $\Delta_0=0.5$, $\Gamma=0.1$ or $0.2$ and $\sigma=0.16$.
This elementary example perfectly illustrates our assertions.
Figure \ref{fake_inversion} corresponds to the case of a false inversion of population, beginning with the state $|1,w(0),z(0)\rangle$.
$\vert \frac{c_2}{c_1}\vert$ becomes very large when $s>0.8$, although $\vert \frac{d_2}{d_1}\vert$
remains very small for the duration of the interaction (using the definition which takes the geometric phases into account to compensate for the unstable norm of the basis vectors). This is consistent with our analysis because this false inversion of populations occurs when $w(s)$ becomes very small due to the gaussian function. Moreover we note the uncontrolled increase in $c_1$ and $c_2$, both reaching values larger than $1$.

Figure \ref{normes} shows the principal cause of this problem (the uncontrollable variations in the norms of $|1,w,z\rangle$ and $|2,w,z\rangle$) and also the exact compensation obtained with the exponential of the geometric phases, leading to two stable unitary norms (since we have taken the precaution of setting the initial norms to one).

In contrast, figure \ref{fake_adiabaticity} illustrates the inverse phenomenon of false adiabaticity. The initial state is $|2,w(0),z(0)\rangle$. Thus the ratio $\vert \frac{c_1}{c_2}\vert$ stays under the value of $0.01$ as if it were an adiabatic evolution, while $\vert \frac{d_1}{d_2}\vert$ increases, with two components of the same order of magnitude ($d_1\simeq d_2$) at the end of the interval.

%%%%%%%%%%%%%%%%%%%%%
% RAJOUT
\subsection{Closed loop around an exceptional point in the parameter space}

If $\Gamma$ and $\phi$ are fixed, the matrix of (\ref{matrice}) has two exceptional points in the parameter plane $(\Omega,\Delta)$. They are located at $(\Omega,\Delta)=(\pm \Gamma/4,0)$. At these points the two eigenvalues and the two eigenvectors coalesce. If a closed loop that encircles a single exceptional point is followed in the parameter space, then the adiabatic basis is transported so as to have the following property~\cite{mailybaev}: 
\begin{eqnarray}
 \vert 1 (s=1) \rangle &=& \nu_1 \vert 2 (s=0) \rangle \nonumber\\
\vert 2 (s=1) \rangle &=& \nu_2 \vert 1 (s=0) \rangle 
\end{eqnarray}
where $\nu_1$ and $\nu_2$ are complex numbers.
If the dynamics is adiabatic, starting with one of the eigenstates $\vert 1(s=0)\rangle$ leads to a final state $\vert 1(s=1)\rangle $ which is proportional to $\vert 2(s=0) \rangle$. This interchange is called an adiabatic flip.
This interesting form of degeneracy in non-hermitian quantum problems is the subject of several recent papers, involving both theoretical \cite{uzdin,berry,gilary2,lefebvre,atabek2} and experimental %on a microwave billiard 
\cite{dietz,dembowski}  studies.
The references \cite{uzdin,gilary2} deal with a symmetric matrix and show that beginning the loop with one state is favourable to an adiabatic behaviour which leads to an adiabatic flip (interchange), while beginning with the other one (the more dissipative one) induces strong non-adiabatic couplings and state exchange during the process, whatever the time duration. The occupancy coefficients of the two states and the states themselves are exchanged during the loop, leading finally to an absence of flip. This is consistent with the conditions associated with the applicability of the adiabatic theorem \cite{nenciurasche,joye}. We can now confirm these results by our calculations and also extend them to the non-symmetric case by using the definition (\ref{gooddefinition}).

\subsubsection{Symmetric case}

\begin{figure}[htp]
\centering
 \includegraphics[width=0.6\linewidth]{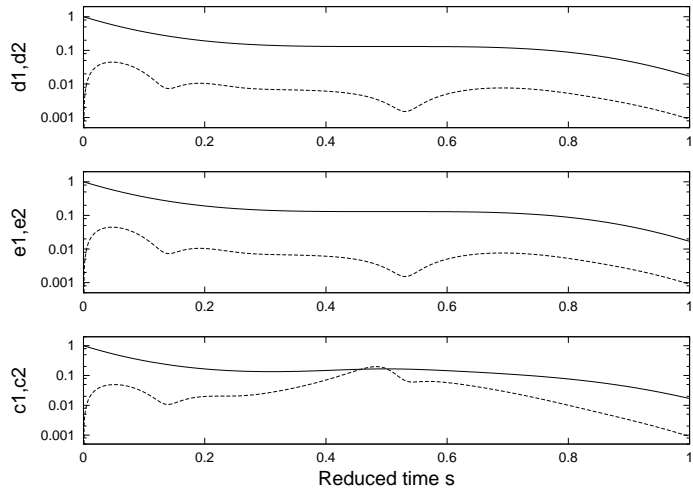}
\caption{Top: $|d_1(s)|$ (line), $|d_2(s)|$ (dashes); middle: $|e_1(s)|$ (line), $|e_2(s)|$ (dashes); bottom: $|c_1(s)|$ (line), $|c_2(s)|$ (dashes). The initial state is $|1,w(0),z(0)\rangle$. }
\label{etat1_sym}
\end{figure}

\begin{figure}[htp]
\centering
 \includegraphics[width=0.6\linewidth]{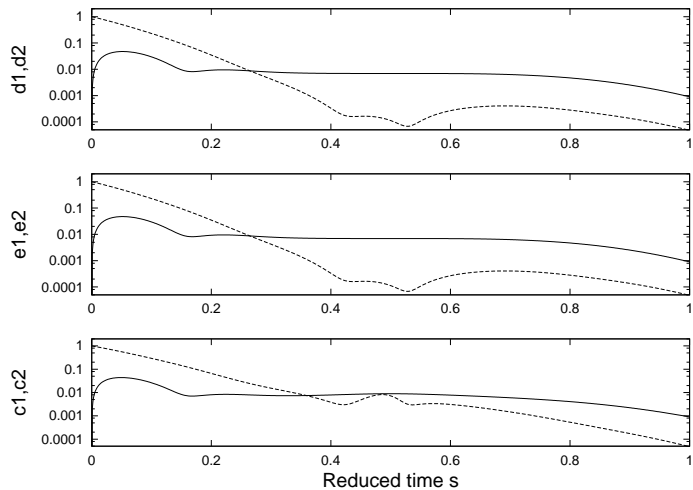}
\caption{Same as figure \ref{etat1_sym} with initial state $|2,w(0),z(0)\rangle$.}
\label{etat2_sym}
\end{figure}

We choose the parameters
\begin{eqnarray}
&&T=100, \nonumber \\ 
&&z(s)=\Delta - \rmi \Gamma /4 \nonumber \\
&&\text{with } \Gamma = 0.5 \text{ and }\Delta(s)= 0.24 \; \Gamma \sin (2\pi s), \nonumber \\
&&w(s)=\Omega(s) = \frac{\Gamma}{4} + 0.24 \; \Gamma \cos (2\pi s).
\end{eqnarray}
We calculate the different adiabatic multipliers $c_1(s),c_2(s)$ and $d_1(s),d_2(s)$ following the adiabatic eigenstates by using the two conventions defined above. %(section \ref{falseadiabaticitysubsection}).
We also calculated results using a third convention which corresponds to the c-product normalization (\ref{cproduct}) and which gives the coefficients 
%\begin{subequations}
\begin{eqnarray}
e_1 &=& \overline{ \langle 1,w(s),z(s)_{\text{c.p.}}} \vert \psi(sT) \rangle, \\ 
e_2 &=&\overline{ \langle 1,w(s),z(s)_{\text{c.p.}}} \vert \psi(sT) \rangle
\label{e1e2}
\end{eqnarray}
%\end{subequations}
such that
\begin{equation}
 \vert \psi(sT) \rangle = e_1(s) \vert 1,w(s),z(s)_{\text{c.p.}} \rangle + e_2(s) \vert 2,w(s),z(s)_{\text{c.p.}} \rangle.
\end{equation}
This c-product normalization is adjusted so as to begin with the same initial condition $(e_1,e_2)=(1,0)$ (or $(e_1,e_2)=(0,1)$) as the two others. 
Figures \ref{etat1_sym} and \ref{etat2_sym} correspond to the dynamics issuing from the initial states $|1,w(0),z(0)\rangle$ and $|2,w(0),z(0)\rangle$, respectively. The $|c_1|$ and $|c_2|$ curves seem to indicate a non-adiabatic behaviour but are not numerically significant. The evolution of $e_1$ and $e_2$ is exactly the same as the evolution of $d_1$ and $d_2$. We see an adiabatic evolution in figure \ref{etat1_sym} for the dynamics issuing from $\vert 1,w(0),z(0)\rangle$. The final state $\vert 1,w(1),z(1)\rangle$ is predominantly occupied at the end of the loop but owing to the flip of the eigenstates we do have an adiabatic flip.
When the initial state is $\vert 2,w(0),z(0)\rangle$ (figure \ref{etat2_sym}), the non-adiabatic exchange of populations during the first part of the loop induces a final state near to state $\vert 1,w(1),z(1)\rangle \propto \vert 2,w(0),z(0) \rangle$ and there is no flip.

% fig 5 etat1_sym.eps
% fig 6 etat2_sym.eps

\subsubsection{Non-symmetric case}

\begin{figure}[htp]
\centering
 \includegraphics[width=0.6\linewidth]{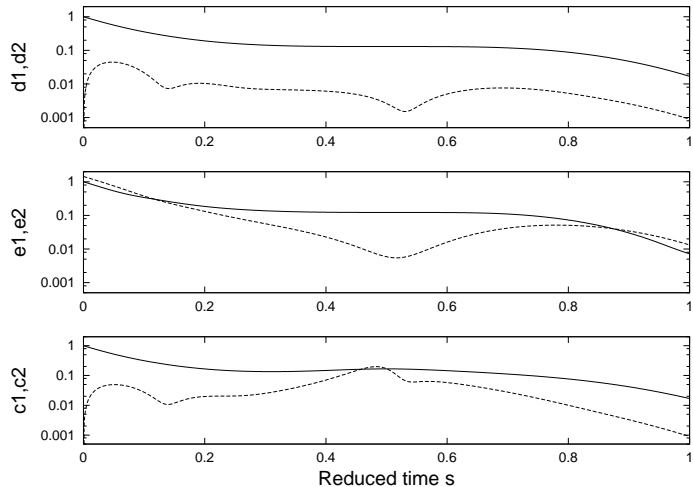}
\caption{Top: $|d_1(s)|$ (line), $|d_2(s)|$ (dashes); middle: $|e_1(s)|$ (line), $|e_2(s)|$ (dashes); bottom: $|c_1(s)|$ (line), $|c_2(s)|$ (dashes). The initial state is $|1,w(0),z(0)\rangle$. }
\label{etat1_nonsym}
\end{figure}

\begin{figure}[htp]
\centering
 \includegraphics[width=0.6\linewidth]{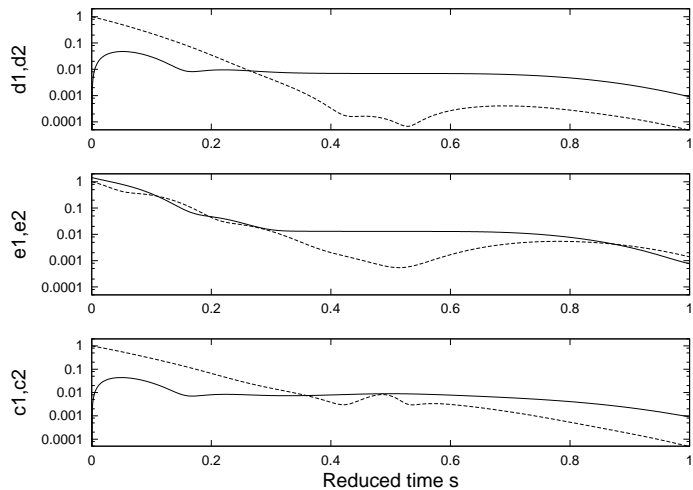}
\caption{Same as figure \ref{etat1_nonsym} with initial state $|2,w(0),z(0)\rangle$.}
\label{etat2_nonsym}
\end{figure}

The adiabatic loop is the same but the only difference from the previous case is that the off-diagonal elements are now complex conjugates, so that the left eigenvectors are not the complex conjugate of the right eigenvectors. We set
\begin{equation}
w(s)=\Omega(s)\, \rme^{\rmi \phi} = \left( \frac{\Gamma}{4} + 0.24 \; \Gamma \cos (2\pi s)\right) \rme^{\rmi\frac{\pi}{4}}
\end{equation}
%The exceptional point is unchanged.
Figures \ref{etat1_nonsym} and \ref{etat2_nonsym} show the evolution of the occupancy coefficients with the different definitions, beginning respectively with the initial state $\vert 1,w(0),z(0) \rangle$ and $\vert 2,w(0),z(0) \rangle$.
We clearly see that it is no more possible to use a c-product-type normalization:
% to begin with the initial condition $(e_1,e_2)=(1,0)$ or $(e_1,e_2)=(0,1)$. 
(\ref{e1e2}) should not be used in this case. The coefficients $c_1$ and $c_2$ continue not to be numerically significant. Only the coefficients $d_1$ and $d_2$ can be used in the present case and they show the same behaviour as for the symmetric case.

%fig 7 etat1_nonsym
%fig 8 etat2_nonsym

\section{Conclusion}

This paper should be regarded as a simple clarification % warning
about the difficult %uncritical 
interpretation of calculations on
dissipative quantum systems described by non-hermitian hamiltonians and about the
use of state populations as important observables,
especially for studies on adiabatic phenomena.
When one works with dissipative quantum systems, the erratic time evolution of the norm of the
adiabatic eigenvectors can create difficulties in answering the question:
what is the relevant definition of a population? 
When the eigenvectors are calculated numerically (for large matrices), there is no reason for them to be continuously transformed from one point to another in time, thus the norm variations can appear to be quite disordered (worse than in our semi-analytic example).

Seemingly, the easiest intuitive solution is to artificialy normalize the ``right'' eigenvectors as if
we were working with an orthogonal basis set, but this is not a coherent way to work.
We have shown that it is much preferable to compensate for the erratic variations in the
norm of the adiabatic basis set by including the
exponentials of the geometric phases in the basis vector decomposition, 
leading to an invariant definition of the populations
under arbitrary changes of the normalization choice.\\
%To avoid any problems, we think that it is preferable to observe the populations themselves instead of a ratio of populations, which could mask an incorrect result due to badly defined adiabatic populations.\\

\ack
%\begin{acknowledgments}
We acknowledge the support from the French Agence Nationale de la Recherche (Project CoMoC). We thank John P. Killingbeck for helping us to clarify the English of the manuscript.
%\end{acknowledgments}


\section*{References}

\begin{thebibliography}{28}
\bibitem{messiah} Messiah A 1959 {\it Quantum Mechanics} (Paris: Dunod)
\bibitem{nenciu} Nenciu G 1980 {\it J. Phys. A: Math. Gen.} {\bf 13} L15
\bibitem{moiseyev} Moiseyev N 2011 {\it Non-Hermitian Quantum Mechanics} (Cambridge: Cambridge University Press)
\bibitem{mailybaev} Mailybaev A A, Kirillov O N and Seyranian A P 2005 {\it Phys. Rev.} A {\bf 72} 014104
\bibitem{viennot} Viennot D, Jolicard G and Killingbeck J P 2008 {\it J. Phys. A: Math. Gen.} {\bf 41} 145303
\bibitem{viennot2} Viennot D, 2009 {\it J. Math. Phys.} {\bf 50} 052101
\bibitem{dridi} Dridi G, Gu\'erin S, Jauslin H R, Viennot D and Jolicard G 2010 {\it Phys. Rev.} A {\bf 82} 022109
\bibitem{moiseyev2} Moiseyev N 1998 {\it Physics Reports} {\bf 302} 211
\bibitem{jolicard} Jolicard G and Austin E J, 1986 {\it Chem. Phys.} {\bf 103} 295
\bibitem{he} He X, Atabek O and Giusti-Suzor A 1988 {\it Phys. Rev.} A {\bf 38} 5586
\bibitem{ben} Ben-Tal N, Moiseyev N, Leforestier C and Kosloff R 1991 {\it J. Chem. Phys} {\bf 94} 7311
\bibitem{atabek} Atabek O, Lefebvre R, Lefebvre C and Nguyen-Dang T T 2008 {\it Phys. Rev.} A {\bf 77} 043413
\bibitem{reviewgeorges1} Killingbeck J P and Jolicard G 2003 {\it J. Phys. A: Math. Gen.} {\bf 36} R105
\bibitem{reviewguerin} Gu\'erin S and Jauslin H R 2003 {\it Adv. Chem. Phys.} {\bf 125} 147
\bibitem{kylstra} Kylstra N J and Joachain C J 1998 {\it Phys. Rev.} A {bf 57} 412
\bibitem{magunov} Magunov A I, Rotter I and Strakhova S I 1999 {\it J. Phys. B: At. Mol. Opt. Phys.} {\bf 32} 1489
\bibitem{gryzlova} Gryzlova E B, Magunov A I, Rotter I and Strakhova S I 2005 {\it Las. Phys.} {\bf 15} 1568
\bibitem{sarandy} Sarandy M S and Lidar D A 2006 {\it Phys. Rev.} A {\bf 73} 062101
\bibitem{yi} Yi X X, Tong D M, Kwek L C and Oh C H 2007 {\it J. Phys. B: At. Mol. Opt. Phys.} {\bf 40} 281
\bibitem{nenciurasche} Nenciu G and Rasche G 1992 {\it J. Phys. A: Math. Gen.} {\bf 25} 5741
\bibitem{joye} Joye A 2007 {\it Comm. Math. Phys.} {\bf 275} 139
\bibitem{uzdin} Uzdin R, Mailybaev A and Moiseyev N 2011 {\it J. Phys. A: Math. Gen.} {\bf 44} 435302
\bibitem{berry} Berry M V and Uzdin R 2011 {\it J. Phys. A: Math. Gen.} {\bf 44} 435303
\bibitem{gilary} Gilary I, Fleisher A and Moiseyev N 2005 {\it Phys. Rev.} A {\bf 72} 012117
\bibitem{gilary2} Gilary I and Moiseyev N 2012 {\it J. Phys. B: At. Mol. Opt. Phys.} {\bf 45} 051002
\bibitem{lefebvre} Lefebvre R, Atabek O, Sindelka M and Moiseyev N 2009 {\it Phys. Rev. Lett.} {\bf 103} 123003
\bibitem{atabek2} Atabek O, Lefebvre R, Lepers M, Jaouadi A, Dulieu O and Kokoouline V 2011 {\it Phys. Rev. Lett.} {\bf 106} 173002
\bibitem{dietz} Dietz B, Harney H L, Kirillov O N, Miski-Oglu M, Richter A and Sch\"afer F 2011 {\it Phys. Rev. Lett.} {\bf 106} 150403
\bibitem{dembowski} Dembowski C, Gr\"af H D, Harney H L, Heine A, Heiss W D, Rehfeld H and Richter A 2001 {\it Phys. Rev. Lett.} {\bf 86} 787
%\bibitem{vitanov} N.V. Vitanov and S. Stenholm, Phys. Rev. A {\bf 55}, 2982 (1997).
%\bibitem{fleisher} A. Fleisher and N. Moiseyev, Phys. Rev. A {\bf 72}, 032103 (2005).
%\bibitem{bohm} A. Bohm, Phys. Rev. A {\bf 60}, 861 (1999).
%\bibitem{madrid} R. de la Madrid, Eur. J. Phys {\bf 26}, 287 (2005).
%\bibitem{madrid2} R. de la Madrid, J. Phys. A {\bf 39}, 9255 (2006)

\end{thebibliography}
\end{document}